\documentclass{endm}
\usepackage{endmmacro}
\usepackage{graphicx}
\usepackage{amsmath}
\usepackage{empheq}
\usepackage{algorithmicx}
\usepackage{algpseudocode}
\usepackage{algorithm}
\usepackage{amsfonts}
\usepackage{array}
\usepackage{caption}
\newtheorem{rmk}{Remark}[section]

\begin{document}

\begin{verbatim}\end{verbatim}\vspace{2.5cm}

\begin{frontmatter}

\title{Formulations for designing robust networks. An application to wind power collection.}

\author{C\'edric Bentz\thanksref{myemail} \thanksref{coemail1}}
\address{CEDRIC-CNAM\\ Paris, France}

\author{Marie-Christine Costa, Pierre-Louis Poirion, Thomas Ridremont\thanksref{coemail2}}
\vspace{-0.5cm}
\address{ENSTA ParisTech and CEDRIC-CNAM\\ Paris, France}



\thanks[myemail]{Thanks to the Gaspard Monge program for optimisation and operational research.}
\thanks[coemail1]{Email: \href{mailto:thomas.ridremont@cnam.fr} {\texttt{\normalshape {(cedric.bentz,thomas.ridremont)}@cnam.fr}}} 
\thanks[coemail2]{Email: \href{mailto:marie-christine.costa@ensta-paristech.fr} {\texttt{\normalshape  {(marie-christine.costa,pierre-louis.poirion)}@ensta-paristech.fr}}} 


\begin{abstract}
	We are interested in the design of survivable capacitated rooted Steiner networks. Given a graph $G=(V,E)$, capacity and cost functions on $E$, a root $r$, a subset $T$ of $V$ of terminals and an integer $k$, we search for a  minimum cost subset $E'\subset E$, covering $T$ and $r$, such that the network induced by $E'$ is $(k+1)$-survivable: after the removal of any $k$ edges, there still exists a feasible flow from $r$ to $T$. We also consider the possibility of protecting a given number of edges. We propose three different formulations: a cut-set, a flow and a bi-level formulation where the second-level is a min-max problem (with an attacker and a defender). We propose algorithms for each problem formulation and compare their efficiency.
\end{abstract}


\end{frontmatter}

\section{Introduction}
\vspace{-0.05cm}
Nowadays, the design of networks is crucial in many fields such as transport, telecommunications or energy. We are interested here in the design of survivable networks \cite{kerivin2005design,magnanti2005strong} able to route a given flow from a root to a set of terminals while respecting capacity constraints. We say that a network is $(k+1)$-survivable if, after any breakdowns on $k$ arcs, it is possible to route the same amount of flow. We also take into account the possibility of protecting a subset of $k'$ arcs of the graph: those arcs cannot be deleted. The resolution of capacitated spanning and steiner tree problems has been studied in \cite{jothi2005approximation,Bentz2016EdgeCapacitated}.

We introduce here the Capacitated Protected Rooted Survivable Network Problem ($\mathbf{CPRSNP}$). Given a directed graph $G=(V,A,c,u)$ where $c$ and $u$ are respectively the cost and the capacity functions on the set of arcs $A$, a set $T \subseteq V$ of terminals, a root $r \in V \setminus T$  and two positive integers $k$ and $k'$ with $k + k' \leq |A|$, $\mathbf{CPRSNP}$ is to find a subset $A' \subseteq A$ of selected arcs of minimum cost and a subset $A'_{p} \subset A'$ with $|A'_{p}| \leq k'$ of protected arcs such that there is a feasible flow (i.e. it respects the arc capacities) routing a unit of flow from $r$ to each vertex of $T$ in the subgraph of $G$ induced by $A'$, even if a breakdown occurs on any $k$ arcs in $A' \setminus A'_{p}$. 

We focus on wiring networks in windfarms, designed to route the energy produced by the wind turbines to the sub-station with respect to some technical constraints (cable capacities, non-splitting constraints, etc.) \cite{hertz2012optimizing,Fischetti2016}.
 Furthermore, we want those networks to be resilient to cable failures. The wind turbines are identical so we can assume that each one produces one unit of energy. Then $A$ is the set of all possible cable locations and $V=T  \cup J \cup \{r\}$ where $T$ (resp $J$) is  the set of terminal nodes (resp. junction nodes between cables), and $r \in V$ is the substation collecting the energy and delivering it to the electric distribution network. In that case, the flow is routed from $T$ to $r$.  In the case of windfarm wiring, one can see the protected arcs as robust cables or additional parallel cables. We consider a budget of protection such that we can only protect $k'$ arcs.

We assume without loss of generality that $u$ is an integer function. In this paper, we work on the oriented version of the problem, however the results can easily be adapted to the non-oriented one.

\section{Formulations of the problem}

\subsection{Definitions and notations}


We add to the input graph a vertex $s$ connected to every terminal $t \in T$ by a fictive arc $(t,s)$ with $c_{ts} = 0$ and $u_{ts} = 1$. Then, $s$ is added to $V$ and the fictive arcs are added to $A$, and we denote by $A_I$ the set of initial arcs. Finding a flow which routes one unit of flow between $r$ and each terminal in the input graph is equivalent to finding a flow of value $|T|$ from $r$ to $s$ in the transformed graph. For each subset $S \subset V$, let $\delta^-(S)$ be the set of arcs entering $S$. We have $\delta^-(S)=\{(i,j) \in A \ | \ i \in V \setminus S ; \ j \in S\}$. We call $\Gamma^-(i)$ and $\Gamma^+(i)$ the sets of respectively predecessors and successors of a vertex $i \in V$.

\vspace{-0.2cm}

\subsection{Cut-set formulation} \label{subsec:cut-set}

We introduce, for all $(i,j) \in A$, a binary variable $y_{ij}$ equal to 1 if and only if the arc $(i,j) \in A'$,  and a binary variable $p_{ij}$ equal to 1 if and only if the arc $(i,j) \in A'_p$. Consider the $r-s$ cuts $[V \setminus V_S,V_S] $ with $ V_S \subset V $, $r \in V \setminus V_S$, $s \in V_S$ and $V_S \neq \{s\}$, and  let $\mathcal{S}$ be the set of all the associated cut-sets in $A$, i.e. $S = \delta^{-}(V_S)$. Notice that if $S \in \mathcal{S}$ then $S \cap A'$ is a cut-set in the selected network.
 For any set $S \subset \mathcal{S}$, let $C_k^S$ be the set of subsets of $S$ of size  $k$.
We define $M_{S}$ as the maximum capacity of a subset of $k$ selected and non-protected arcs of $S$: 

\begin{subequations}
	\begin{empheq}{align}
	M_{S} \quad = \quad \max\limits_{C \in C^S_k} \sum\limits_{(i,j) \in C}  u_{ij}(y_{ij}-p_{ij}) \label{eq:MS}
	\end{empheq}
\end{subequations}

$M_S$ corresponds to the maximum capacity that can be lost in the cut-set $S$ after the deletion of $k$ non-protected arcs. We propose the following cutset formulation:
\vspace{-0.3cm}

\begin{subequations}
	\begin{empheq}{align}
	\min\limits_{y,p}  & \  \sum\limits_{(i,j) \in A} c_{ij}y_{ij} \quad \notag \\[-3pt]
	\text{s.t.}  & \ \hspace{-.3cm}\sum\limits_{(i,j) \in \delta^-(V_S)}\hspace{-.3cm} u_{ij}y_{ij} \ - \ M_{S} \ \geq \ |T| & \quad \forall S \in \mathcal{S} \quad \label{cstr:cutset-cut}\\[-3pt]
	& \ \sum\limits_{(i,j) \in A}  p_{ij} \quad \leq \quad k' & \label{cstr:cutset-prot}\\[-3pt]
	& \ \quad y_{ts} \quad = \quad 1 \qquad & \forall t \in T \quad \label{cstr:cutset-yfict}\\[-3pt]
	& \ \quad p_{ts} \quad = \quad 0 \qquad & \forall t \in T \quad \label{cstr:cutset-pfict}\\[-3pt]
	&  \ \quad y_{ij},p_{ij} \in \{0,1\} \qquad & \forall (i,j) \in A \quad \notag
	\end{empheq}
\end{subequations}

Constraints (\ref{cstr:cutset-cut}) ensure that for each cut, the capacity of the cut after the worst-case deletion of $k$ non-protected arcs of the cut-set is at least equal to the number of terminals, i.e. the graph is $(k+1)$-survivable. 
Constraints (\ref{cstr:cutset-prot}) ensure that we have at most $k'$ protected arcs while (\ref{cstr:cutset-yfict}) and (\ref{cstr:cutset-pfict}) guarantee that all fictive arcs are selected and none of them are protected. Notice that we do not have to impose $p \leq y$ since, from any optimal solution, we can get an equivalent one where this holds.

Constraints (\ref{cstr:cutset-cut}) are non linear because of the use of the maximum operator in the definition of $M_{S}$. To linearize it, we add the following constraints:
\vspace{-.15cm}
\begin{subequations}
	\begin{empheq}{align}
	M_{S} \quad \geq \quad \sum_{(i,j) \in C} u_{ij}(y_{ij}-p_{ij}) \quad \forall S \in \mathcal{S}, \ \forall C \in C^S_k \label{cstr:cutset-MS}
	\end{empheq}
\end{subequations}

The number of constraints (\ref{cstr:cutset-cut}) and (\ref{cstr:cutset-MS}) being exponential (and so is the number of variables $M_{S}$), we propose a constraints-and-columns generation algorithm. We begin with a small subset of $\mathcal{S}$ in constraints (\ref{cstr:cutset-cut}) and (\ref{cstr:cutset-MS}). We obtain a lower bound for our problem. Then we select a cut-set that does not verify some constraint (\ref{cstr:cutset-cut}): given a network induced by the current value of $y$, we find by a using a MIP the cut of minimum residual capacity once we delete its $k$ most capacitated non-protected arcs. If this capacity is inferior to $|T|$, we add the constraints associated to this cut-set, otherwise the algorithm terminates.

\subsection{Flow formulation}\label{subsec:flowForm}

Now we define $\mathcal{F}$ as the set of all possible arc-failure scenarios: it corresponds to the set of all $k$-combinations in $A_I$. We introduce the variable $x_{ij}^{F}$ which represents the amount of flow routed through the arc $(i,j) \in A$ when the scenario $F \in \mathcal{F}$ occurs. The variables $y$ and $p$ are defined as in the previous formulation (Section \ref{subsec:cut-set}). We propose the following flow formulation:

\vspace{-.15cm}
\begin{subequations}
	\begin{empheq}{align}
	\min\limits_{x,y,p}   &  \sum\limits_{(i,j) \in A} c_{ij}y_{ij} \quad \notag\\[-3pt]
	\text{s.t. }  & \hspace{-.2cm}\sum\limits_{i \in \Gamma^-(j)} \hspace{-.2cm} x^{F}_{ij} \ - \ \hspace{-.3cm}\sum\limits_{k \in \Gamma^+(j)} \hspace{-.2cm}x^{F}_{jk} = 0  & \forall j \in V \setminus \{r,s\}, \ \forall F \in \mathcal{F} \label{cstr:flow-flowCons} \\[-3pt]
	& \hspace{-.2cm}\sum\limits_{t \in \Gamma^-(s)}\hspace{-.2cm} x^{F}_{ts} \quad  =  \quad |T|  &\forall F \in \mathcal{F} \label{cstr:flow-flowConsSink}\\
	& \hspace{-.2cm}\sum\limits_{(i,j) \in A_I}\hspace{-.2cm} p_{ij} \quad  \leq  \quad k' & \label{cstr:flow-protCons}\\[-3pt]
	& x^{F}_{ij} \quad  \leq  \quad u_{ij}y_{ij}  & \forall (i,j) \in A, \ \forall F \in \mathcal{F} \label{cstr:flow-cap}\\[-3pt]
	& x^{F}_{ij} \quad \leq \quad u_{ij}p_{ij}  & \forall F \in \mathcal{F}, \ \forall (i,j) \in F \label{cstr:flow-deletedArc}\\[-3pt]
	&  \omit\rlap{$x \in \mathbb{R}_+^{|A| \times |\mathcal{F}|}, \quad y \in \{0,1\}^{|A|}, \quad p \in \{0,1\}^{|A_I|} $} & \notag
	\end{empheq}
\end{subequations}

Constraints (\ref{cstr:flow-flowCons}) and (\ref{cstr:flow-flowConsSink}) ensure that there is a flow of value $|T|$ for each arc-failure scenario. Constraint (\ref{cstr:flow-protCons}) ensure that we have at most $k'$ protected arcs. Constraints (\ref{cstr:flow-cap}) are the capacity constraints, for each scenario of failure. Finally, constraints (\ref{cstr:flow-deletedArc}) ensure that in a scenario $F$ where $(i,j) \in F$, we can route some flow through $(i,j)$ only if the arc is protected. Notice that w.l.o.g. we can assume that $p \leq y$. 

The number of variables $x^{F}_{ij}$ and constraints (\ref{cstr:flow-flowCons}) and (\ref{cstr:flow-flowConsSink})  being exponential, as in the previous section we propose a constraints-and-columns generation algorithm to solve the problem. We begin with a small subset of $\mathcal{F}$. The separation problem is the problem of the $k$ most vital links in a flow network \cite{ratliff1975finding}: we search for the $k$ non-protected arcs which, once simultaneously deleted, reduce the most the value of the maximum $s-t$ flow.

\vspace{-0.3cm}

\subsection{Bilevel formulation}

The bilevel formulation proposed here is particular in that the second-level is a $\min \max$ problem. It can be seen as a game with a defender and an attacker (corresponding respectively to the leader and the follower).

For each $(i,j) \in A$, we introduce a variable $x_{ij}$ which corresponds to the amount of flow that the defender chooses to route through the arc $(i,j)$. The variables $y$ and $p$ are defined as in Section \ref{subsec:cut-set}. We also introduce the binary variables  $b_{ij}$, $\forall (i,j) \in A$:  $b_{ij}= 1$ if and only if the attacker chooses to delete the arc $(i,j)$. We define the following polyhedron: 
\vspace{-.15cm}
\[
\mathcal{X}(y,b,p)=\left\{
x \in \mathbb{R}^{|A|}_+ \left|
\begin{array}{ll}
\sum\limits_{i \in \Gamma^-(j)} \hspace{-.3cm} x_{ij} -  \hspace{-.3cm}\sum\limits_{k \in \Gamma^+(j)} \hspace{-.3cm} x_{jk} = 0 &\forall j \in V \setminus \{r,s\}\\[-4pt]
x_{ij} \ \leq \ u_{ij}y_{ij} \ & \forall (i,j) \in A\\[-4pt]
x_{ij} \ \leq \ u_{ij}(1 - b_{ij} + p_{ij})  \ & \forall (i,j) \in A_I
\end{array}
\right.
\right\}
\]

This polyhedron $\mathcal{X}(y,b,p)$ corresponds to the set of possible flows on the subgraph of $G$ induced by the arcs $(i,j)$ such that $y_{ij}=1$ provided they have not been deleted (thus non-protected). In $\mathcal{X}(y,b,p)$ there are the flow conservation constraints, the capacity constraints and the constraints imposing a flow equal to 0 on any arc which is non-protected and deleted. We also define the two following polyhedrons: 

\begin{center}
$\mathcal{B} = \{ b \in \{0,1\}^A \text{ } | \sum_{(i,j) \in A}  b_{ij} \leq k \text{ } ; \text{ } b_{ts} = 0 \text{ } \forall t \in T\}$

$ \mathcal{Y} = \{(y,p)\in\{0,1\}^{A^2} \text{ } | \ p \leq y \ ; \ \hspace{-.2cm} \sum\limits_{(i,j) \in A} \hspace{-.2cm} p_{ij} \leq k' \}$
\end{center}

The polyhedron $\mathcal{B}$ defines the set of possible scenarios of arc failures while $\mathcal{Y}$ defines the set of possible selected and protected arcs (there are at most $k'$ protected arcs in the selected
 network). Then we propose the following bilevel program:
\vspace{-0.2cm}
\begin{subequations}
	\begin{empheq}{align}
	\min\limits_{(y,p) \in \mathcal{Y}} \quad &  \sum\limits_{(i,j) \in A} c_{ij}y_{ij} &&&& \notag\\[-3pt]
	\text{s.t.} \quad & f(y,p) \geq |T| &&&& \label{fYCstr}\\[-3pt]
	&\text{where } f(y,p) = & \hspace{-.2cm} \min\limits_{b \in \mathcal{B}} \quad & \max\limits_{x \in \mathcal{X}(y,b,p)} &&  \hspace{-.3cm} \sum\limits_{j \in \Gamma^+(r)} x_{rj}&&
	\end{empheq}
\end{subequations}

At the upper-level, the defender selects the set of arcs he wants to add to the network as well as the one he wants to protect, by choosing a couple $(y,p)$ in $\mathcal{Y}$. The attacker then deletes some arcs by setting the variable $b \in \mathcal{B}$ in order to minimize the maximum flow that the defender will compute by setting the variable $x$ in the flow polyhedron $\mathcal{X}(y,b,p)$. The aim of the defender is to ensure that this flow is at least equal to $|T|$ (Constraint (\ref{fYCstr})).\\

Consider the $\max$ problem in the lower-level: $y$, $p$ and $b$ are already fixed to $\hat{b}$, $\hat{p}$ and $\hat{y}$. The problem is a max-flow problem from $r$ to $s$, with two sets of capacity constraints. In our problem, the flow must be integer since it corresponds to a number of terminals. However, it is well known that the matrix of coefficients $M$ in the arc-formulation of a max-flow is totally unimodular. Then, adding the second set of capacity constraints is equivalent to appending the identity matrix to $M$: the matrix remains totally unimodular and the capacities being integer, we ensure that the extreme points of the polyhedron defined by $\mathcal{X}(y,b,p)$ are integer. Thus, we can relax integrality constraints on $x$.\\

There always exists a feasible flow of value 0 and the lower-level problem is also trivially upper bounded. Hence, the strong duality holds and we can introduce the dual of the lower-level problem, after a slight reformulation due to the totally unimodular matrix: 
\vspace{-0.4cm}
\begin{subequations}
	\begin{empheq}{align}
	\min\limits_{\lambda, \mu, \gamma} & \quad \rlap{$\sum\limits_{(i,j) \in A} u_{ij}\hat{y}_{ij}\lambda_{ij} + u_{ij}(1 - \hat{b}_{ij} + \hat{p}_{ij})\gamma_{ij}$} \notag\\[-4pt]
	\text{s.t} & \quad \lambda_{ij} + \gamma_{ij} - \mu_{i} + \mu_{j} & \geq \quad & 0 \quad & \forall (i,j) \in A \label{lambdaGammaCstr}\\[-4pt]
	& \quad \mu_{r} & =  \quad& 1 \quad & \label{muRConstraints}\\[-4pt]
	& \quad \mu_{s} & =  \quad& 0 \quad & \label{muSConstraints}\\[-4pt]
	& \quad \rlap{$\lambda,\gamma \ \in \ [0,1]^{|A|}, \qquad \mu \ \in \ [0,1]^{|V|}$} \notag
	\end{empheq}
\end{subequations}
This problem is a special formulation of a min-cut problem: $\mu$ defines the two parts of the cut (sets of vertices $i$ such that either $\mu_i =0$ or $\mu_i =1$). The variable $\gamma_{ij}$ is equal to 1 at least for all the edges $(i,j)$ in the cut-set with $p_{ij} = 0$ and $b_{ij} = 1$, $\lambda$ is equal to 1 for all other edges in the cut-set (if an edge $(u,v)$ is not in the cut-set, we have $\lambda_{uv} =  \gamma_{uv} = 0$). We denote the polyhedron defined by the dual constraints by $\mathcal{D}$.

As the lower-level can be reformulated as a $\min \min$ function by using the dual described above, it can then be rewritten as follows:

\vspace{-0.2cm}

\begin{equation*}(2LP) \left |
\begin{array}{ll}
\min\limits_{b,\lambda, \mu, \gamma} & \quad \sum\limits_{(i,j) \in A} u_{ij}\hat{y}_{ij}\lambda_{ij} + u_{ij}(1 - b_{ij} + \hat{p}_{ij})\gamma_{ij} \notag\\[-4pt]
\text{s.t} \quad & \quad b \in \mathcal{B}\\[-4pt]
& \quad (\lambda, \mu, \gamma) \in \mathcal{D}
\end{array}
\right.
\end{equation*}

At this point, $b$ is a variable so the objective function is non-linear. We linearize the terms $b_{ij} \gamma_{ij}$ in a classical way  by introducing binary variables $l_{ij}$ verifying the set of constraints  $\ l_{ij} \leq b_{ij}$, $\ l_{ij} \leq \gamma_{ij}$ and $\ l_{ij} \geq \gamma_{ij} - (1 - b_{ij}), \quad \forall (i,j) \in A$, which is denoted by $\mathcal{L}(b,\gamma)$.
We also define the function $g(y,p,\lambda,\gamma, l) =  \sum_{(i,j) \in A} u_{ij}y_{ij}\lambda_{ij} + u_{ij}\gamma_{ij} - u_{ij}l_{ij} + u_{ij}p_{ij}\gamma_{ij}$. We can then rewrite the bilevel program as:

\vspace{-0.2cm}
\begin{empheq}{align*}
\min\limits_{(y,p) \in \mathcal{Y}} \quad &  \sum\limits_{(i,j) \in A} c_{ij}y_{ij} &&&& \notag\\[-2pt]
\text{s.t. } \quad & f(y,p) \geq |T| &&&& \label{fYCstr}\notag\\[-2pt]
&\text{where }\quad & f(y,p) \quad = & \min\limits_{b, \lambda, \gamma, \mu, l} \quad & g(y,p,\lambda,\gamma, l) &&&& \notag\\[-2pt]
&  &  \text{s.t. } \quad& b \in \mathcal{B} &&&& \notag\\[-2pt]
&  &  \quad& l \in \mathcal{L}(b,\gamma) &&&& \notag\\[-2pt]
&  &  \quad& (\lambda, \mu, \gamma) \in \mathcal{D} &&&& \notag
\end{empheq}

In this bilevel reformulation, the polyhedron of the lower-level defined by $\mathcal{B}$, $\mathcal{L}(b,\gamma)$ and $\mathcal{D}$  depends on neither $y$ nor $p$. We can then consider the convex hull of the lower-level polyhedron and we denote by $\mathcal{H}$ the set of its extreme points; $(\hat{b}^h, \hat{\lambda}^h, \hat{\gamma}^h, \hat{\mu}^h, \hat{l}^h)$ are respectively the values of $(b, \lambda, \gamma, \mu, l)$ at the extreme point $h \in \mathcal{H}$. We can then reformulate the bilevel formulation as a single-level one as follows:

\vspace{-0.2cm}
\begin{subequations}
	\begin{empheq}{align}
	\min    &  \sum\limits_{(i,j) \in A} c_{ij}y_{ij} &\notag\\\vspace{-0.3cm}
	\text{s.t.} \quad & g(y,p,\hat{\lambda}^h,\hat{\gamma}^h, \hat{l}^h)  \ \geq \ |T| & \forall h \in \mathcal{H}\label{cstr:biGA} \\[-4pt]
	\mathbf{(BP)} \hspace{2cm}&  b \in \mathcal{B} & \label{cstr:polyB}\\[-4pt]
	&  l \in \mathcal{L}(b,\gamma) & \label{cstr:polyL}\\[-4pt]
	&  (y,p) \in \mathcal{Y} & \label{cstr:polyYP}\\[-4pt]
	&  (\lambda, \mu, \gamma) \in \mathcal{D} \label{cstr:polyD}& 
	\end{empheq}
\end{subequations}

Constraints (\ref{cstr:biGA}) ensure that for each extreme point of $\mathcal{H}$, $f(y,p)$ is greater than $|T|$ (i.e. the minimum value of $f(y,p)$ over the polyhedron defined by (\ref{cstr:polyB}), (\ref{cstr:polyL}) and (\ref{cstr:polyD}) is greater than $|T|$), meaning that the flow cannot be decreased after any $k$ breakdowns.

\begin{rmk}
	In $\mathbf{(BP)}$, $g(y,p,\lambda,\gamma, l)$ is non-linear because of the products $y_{ij}\lambda_{ij}$ and $p_{ij}\gamma_{ij}$ but they can be linearized as it has been done for  $b_{ij} \gamma_{ij}$ above.
\end{rmk}

However, there is an exponential number of constraints (\ref{cstr:biGA}) and we do not know how to describe explicitly the convex hull $\mathcal{H}$. To tackle this issue, we use a constraints generation algorithm where we relax the set of constraints (\ref{cstr:biGA})  and use $(2LP)$ as the separation problem: while the optimum value of $(2LP)$ is lower than $|T|$ for the current optimal solution $(\hat{y}, \hat{p})$, we generate the constraint (\ref{cstr:biGA}) associated to the extreme point corresponding to the optimal values of $(b, \lambda, \gamma, \mu, l)$ in $(2LP)$.

\begin{prop}\label{betterCutProp}
	Let $(\hat{y}^{1},\hat{p}^{1})$ and $(\hat{y}^{2},\hat{p}^{2})$ be two feasible solutions of $\mathbf{(BP)}$ such that $\hat{y}^{1} \geq \hat{y}^{2}$ and $\hat{p}^{1} \geq \hat{p}^{2}$. If adding a constraint $g(y,p,\lambda,\gamma,l) \leq g(y,p,\hat{\lambda}^a, \hat{\gamma}^a, \hat{l}^a)$ makes any solution with $(y,p) = (\hat{y}^{1},\hat{p}^{1})$ infeasible, then it also makes any solution with $(y,p) = (\hat{y}^{2},\hat{p}^{2})$ infeasible.
\end{prop}
\vspace{-0.2cm}
\begin{proof}
	For any value $(\hat{\lambda}^a, \hat{\gamma}^a, \hat{l}^a)$ of $(\lambda, \gamma, l)$, we have $g(\hat{y}^{1},\hat{p}^{1},\hat{\lambda}^a, \hat{\gamma}^a, \hat{l}^a) \geq g(\hat{y}^{2},\hat{p}^{2}\hat{\lambda}^a, \hat{\gamma}^a, \hat{l}^a)$ since $\hat{y}^{1} \geq \hat{y}^{2}$ and $\hat{p}^{1} \geq \hat{p}^{2}$ (as $u$, $\lambda$ and $\gamma$ are positive). Hence, if $g(\hat{y}^{1},\hat{p}^{1},\hat{\lambda}^a, \hat{\gamma}^a, \hat{l}^a) \leq |T| - 1$ then  $g(\hat{y}^{2},\hat{p}^{2},\lambda^S,\gamma^S,l^S) \leq |T| - 1$.
\end{proof}
\vspace{-0.1cm}

To improve the cut obtained by solving $(2LP)$, we try to inject better values $\hat y$ of variables $y$ in it. To get these values, we first solve the following problem, and then we compute the new $\hat y$ accordingly (as explained later). Given a starting solution $(\hat{y}, \hat{p})$, we want to find a cut-set in the support network with a minimum number of arcs such that this cut-set is non-valid in the network induced by $(\hat{y},\hat{p})$ (meaning that if we remove $k$ non-protected arcs of the cut-set, its remaining capacity is smaller than $|T|$). This can be modeled as follows:
\vspace{-0.3cm}
\begin{subequations}
	\begin{empheq}{align}
	\min  \quad  &  \sum_{(i,j) \in A} \lambda_{ij} \notag\\[-4pt]
	\text{s.t} \quad& \sum_{(i,j) \in A} u_{ij}\hat{y}_{ij}\lambda_{ij} + \hat{p}_{ij}\gamma_{ij} & \leq & \quad |T| - 1 \label{cstr:enhanceNF}\\[-4pt]
	& \sum_{(i,j) \in A} \gamma_{ij} & \leq & \quad k & \label{cstr:enhanceK}\\[-4pt]
	&\ \gamma_{ts} & = & \quad 0 \quad & \forall t \in T \label{cstr:enhanceFic}\\
	& \omit\rlap{$\quad (\lambda, \mu, \gamma) \in \mathcal{D} $} &&&\notag\\
	& \qquad \mu \in \{0,1\}^V\notag
	\end{empheq}
\end{subequations}

The variables $(\lambda, \mu, \gamma)$ define a cut as in $(2LP)$ since they belong to $\mathcal{D}$ (recall that $\mathcal{D}$ is the set constraints (6) ). However, adding the other constraints makes the constraints matrix not unimodular anymore: we have to set $\mu$ as a 0-1 variable. Constraint (\ref{cstr:enhanceNF}) ensures that the cut-set selected is non-valid (as defined before). Constraint (\ref{cstr:enhanceK}) bounds the number of deleted arcs to at most $k$, while constraints (\ref{cstr:enhanceFic}) forbid the deletion of fictive arcs. 

Then the new values of $\hat y_{ij}$ are computed as follows: we set $\hat y_{ij}$ to 1 for all $(i,j)$ with $\lambda_{ij} = \gamma_{ij} = 0$ and let the others to their current value. It implies that the new value of $y$ will include the previous one and, using Proposition \ref{betterCutProp}, we generate a better constraint than the original by calculating the extreme points associated to this new value of $y$. 

\section{Tests and conclusion}

All experiments were performed on a computer with a 2.40GHz Intel(R) Core(TM) i7-5500U CPU and a 16GB RAM using the solver CPLEX version 12.6.1. We present 3 tables of results. For each table, we present the solving time (limited to 2000 seconds) and the gap if the optimum is not reached. The cut-set and flow formulations were solved using an iterative procedure in which after the generation of the constraint, the model is resolved as a MIP.

Table \ref{tab:compUniform} shows the performances of the different formulation on an uniform capacities generated instance (those results reflect the general tendency obtained on tests made on other instances) with no protected arcs. One can see that the cut-set formulation is the most efficient one: this can be explained by the fact that the variable $M_S$ is a constant in this case. The bilevel formulation is also efficient here whereas the flow formulation becomes inefficient for $k \geq 2$.  \\
 
\captionsetup{font=scriptsize,labelfont=scriptsize}

\begin{tabular}{c}
	\centering
	\begin{minipage}{0.85\linewidth}
		\centering
		\scalebox{0.7}{
			\centering
			\begin{tabular}{|c|c|c|c|c|c|c|c|c|}
				\hline
				\multicolumn{3}{|c|}{Instance} & \multicolumn{2}{c|}{Bilevel} & \multicolumn{2}{c|}{Cut-set} & \multicolumn{2}{c|}{Flow}\\
				\cline{1-9}
				\hline
				$|V|$-$|T|$-$|A|$ & k & k' & t (s) & gap & t (s) & gap & t (s)& gap\\
				\hline		
				20-5-90 & 1 & 0 &5.9& 0 &2.17&0&21.3& 0\\
				- & 2 & 0 &49.3& 0 &7.8&0&2000& 0.25\\
				- & 3 & 0 &22.8& 0 &21.1&0&2000& 0.15\\
				\hline
			\end{tabular}}
			\centering
			\captionof{table}{Comparison of the performance for uniform capacities with $k' = 0$}
			\label{tab:compUniform}	
		\end{minipage} 
	\end{tabular}\\
	
Table \ref{tab:comp3} shows the difference of performance between the three different formulations for different values of $k$ and $k'$ on a generated instance with non-uniform capacities. Although the flow formulation competes with the bilevel one for $k=1$, it seems obvious that the bilevel formulation is the best one to solve $\mathbf{CPRSNP}$ when $k > 1$. This can be explained by the fact that the number of variables ($M_S$ and $x$) and constraints is exponential in $k$ for the two other formulations.

Table \ref{tab:bilevelRes} shows the results obtained on three different instances of the bilevel formulation for different values of $k$ and $k'$. We can see that even if the solving time generally grows with $k$ and $k'$, the formulation is much less sensitive to those two parameters.

	\begin{tabular}{cc}
		\begin{minipage}{.35\linewidth}
			\scalebox{0.6}{\hspace{-1cm}\begin{tabular}{|c|c|c|c|c|c|c|c|}
			\hline
			\multicolumn{2}{|c|}{I} & \multicolumn{2}{c|}{Bilevel} & \multicolumn{2}{c|}{Cut-set} & \multicolumn{2}{c|}{Flow}\\
			\cline{1-8}
			\hline
			k & k' & t (s) & gap & t (s) & gap & t (s)& gap\\
			\hline		
			1 & 0 &9.3& 0 &1812.2&0&15.8& 0\\
			1 & 1 &11.6& 0 &658.8&0&5.14& 0\\
			1 & 2 &13.0& 0 &691.3&0&5.3& 0\\
			2 & 0 &43.8& 0 &2000&0.12&2000&  0.01 \\
			2 & 1 &30.9& 0 &2000&0.22&2000&  0.05 \\
			2 & 2 &30.5& 0 &2000&0.04&112.6& 0\\
			3 & 0 &29.0& 0 &2000&0.19&2000& 0.18 \\
			3 & 1 &21.0& 0 &2000&0.24&2000& 0.21\\
			3 & 2 &66.5& 0 &2000&0.2&2000& 0.16\\
			\hline
		\end{tabular}}
		\captionof{table}{Comparison of the 3 formulations for $|V| = 30, |T| = 3, |A| = 140$}
		\label{tab:comp3}
		\end{minipage} &
		
		\begin{minipage}{.65\linewidth}
			\scalebox{0.6}{\hspace{.5cm}\begin{tabular}{|c|c|c|c|c|c|c|c|c|c|}
					\hline
					\multicolumn{2}{|c|}{Instance} & \multicolumn{2}{c|}{$k' = 0$} & \multicolumn{2}{c|}{$k' = 1$} & \multicolumn{2}{c|}{$k' = 2$} & \multicolumn{2}{c|}{$k' = 3$}\\
					\cline{1-10}
					\hline
					$|V|$-$|T|$-$|A|$ & k & t (s) & gap & t (s) & gap & t (s)& gap & t (s)& gap\\
					\hline
					20-5-100 & 1 &6.4& 0 &19.0 & 0&26.5& 0& 18.0& 0\\
					- & 2 &15.2& 0 &58.9& 0&167.2& 0&223.7& 0\\
					- & 3 & - & - &58.2& 0 &315.8& 0&822.3& 0 \\
					25-8-120 & 1 &22.0& 0 &152.0 & 0&87.9& 0& 67.3& 0\\
					- & 2 &76.4& 0 &161.1& 0&546.3& 0&794.4& 0\\
					- & 3 & - & - &34.2& 0 &158.0& 0&1670.3& 0 \\
					35-3-175 & 1 &35.7& 0 &101.4 & 0&82.3& 0& 159.6& 0\\
					- & 2 &180.6& 0 &1077.3& 0&2000& 0.11&1245.7& 0\\
					- & 3 & 244.3 & 0 &2000& 0.03 &2000& 0.45&2000& 0.49 \\
					\hline
				\end{tabular}}
			\captionof{table}{Performance of the bilevel formulation}
					\label{tab:bilevelRes}
		\end{minipage} 
	\end{tabular}\\

\vspace{-0.2cm}

\bibliographystyle{plain}
\bibliography{BibThese}

\begin{thebibliography}{1}

\bibitem{Bentz2016EdgeCapacitated}
C{\'{e}}dric Bentz, Marie{-}Christine Costa, and Alain Hertz.
\newblock On the edge capacitated steiner tree problem.
\newblock {\em CoRR}, abs/1607.07082:1 -- 31, 2016.

\bibitem{Fischetti2016}
Martina Fischetti and Michele Monaci.
\newblock Proximity search heuristics for wind farm optimal layout.
\newblock {\em Journal of Heuristics}, 22(4):459--474, 2016.

\bibitem{hertz2012optimizing}
Alain Hertz, Odile Marcotte, Asma Mdimagh, Michel Carreau, and Francois Welt.
\newblock Optimizing the design of a wind farm collection network.
\newblock {\em INFOR: Information Systems and Operational Research},
  50(2):95--104, 2012.

\bibitem{jothi2005approximation}
Raja Jothi and Balaji Raghavachari.
\newblock Approximation algorithms for the capacitated minimum spanning tree
  problem and its variants in network design.
\newblock {\em ACM Transactions on Algorithms (TALG)}, 1(2):265--282, 2005.

\bibitem{kerivin2005design}
Herv{\'e} Kerivin and A~Ridha Mahjoub.
\newblock Design of survivable networks: A survey.
\newblock {\em Networks}, 46(1):1--21, 2005.

\bibitem{magnanti2005strong}
Thomas~L Magnanti and S~Raghavan.
\newblock Strong formulations for network design problems with connectivity
  requirements.
\newblock {\em Networks}, 45(2):61--79, 2005.

\bibitem{ratliff1975finding}
H~Donald Ratliff, G~Thomas Sicilia, and SH~Lubore.
\newblock Finding the n most vital links in flow networks.
\newblock {\em Management Science}, 21(5):531--539, 1975.

\end{thebibliography}
\end{document}